\documentclass[aps,prl,superscriptaddress,floatfix,twocolumn,showpacs]{revtex4}

\usepackage{verbatim}

\usepackage{amsmath}
\usepackage{amsfonts}
\usepackage{amssymb}
\usepackage{graphicx}
\usepackage{bm}
\usepackage{color}
\usepackage{epsf}

\usepackage{psfrag}

\def\bea{\begin{eqnarray}}
\def\eea{\end{eqnarray}}
\def\beas{\begin{eqnarray*}}
\def\eeas{\end{eqnarray*}}
\def\beqas{\begin{eqnarray*}}
\def\eqas{\end{eqnarray*}}
\def\beq{\begin{equation}}
\def\eeq{\end{equation}}
\def\beqd{\begin{displaymath}}
\def\eeqd{\end{displaymath}}
\def\eqd{\end{displaymath}}

\def\slashchar#1{\setbox0=\hbox{$#1$}
   \dimen0=\wd0
   \setbox1=\hbox{/} \dimen1=\wd1
   \ifdim\dimen0>\dimen1
      \rlap{\hbox to \dimen0{\hfil/\hfil}}
      #1
   \else\begin{eqnarray}
      \rlap{\hbox to \dimen1{\hfil$#1$\hfil}}
      /
   \fi}

\begin{document}
\title
{Neutrino-production of a charmed meson and the transverse spin structure of the nucleon}
\author{ B.~Pire}
\affiliation{ CPHT, \'Ecole Polytechnique,
CNRS, 91128 Palaiseau,     France }
\author{ L.~Szymanowski}
\affiliation{ National Centre for Nuclear Research (NCBJ), Warsaw, Poland}

\date{\today}

\begin{abstract}

\noindent
We calculate the amplitude for exclusive  neutrino production of a charmed meson on an unpolarized target, in the colinear QCD approach where generalized parton distributions (GPDs) factorize from  perturbatively calculable coefficient functions. We demonstrate that the transversity chiral odd   GPDs contribute to the transverse cross section if the hard amplitude is calculated up to  order $m_c/Q$. We show how to access these GPDs through the azimuthal dependence of the $\nu N \to \mu^- D^+ N$ differential cross section.
\end{abstract}

\maketitle

\paragraph*{Introduction.}
The  transverse spin structure of the nucleon - that is  the way quarks and antiquarks spins share the polarization of a nucleon, when it is polarized transversely to its direction of motion - is almost completely unknown. The transversity distributions which encode this information have proven to be among  the most difficult hadronic quantities to access. This is  due to the chiral odd character of the quark operators which enter their definition; this feature enforces the decoupling of these distributions from  most measurable hard amplitudes. After the pioneering works \cite{trans}, much effort \cite{Barone} has been devoted to the exploration of many channels but experimental difficulties have challenged some of the most promising ones. Some very interesting information has been obtained through  the chiral-odd transverse momentum distribution functions (TMDs) which have been extracted  \cite{TMD,BC} from recent semi-inclusive deep inelastic scattering  data.
 
It is now well established that generalized parton distributions (GPDs) give access to the internal structure of hadrons in a much more detailed way than parton distributions (PDFs) measured in inclusive processes, since they allow a 3-dimensional analysis \cite{3d}. The study of exclusive reactions mediated by a highly virtual photon in the generalized Bjorken regime benefits of the factorization properties of the leading twist QCD amplitudes \cite{fact1,fact2,fact3}
for reactions such as deeply virtual Compton scattering. A welcome feature of this formalism is that spin related quantities such as helicity or transversity GPDs may be accessed in reactions on an {\em unpolarized} nucleon.

Neutrino production is another way to access (generalized) parton distributions  \cite{weakGPD}. Neutrino induced cross sections are orders of magnitudes smaller than those for electroproduction and neutrino beams are much more difficult to handle than charged lepton beams; nevertheless, they have  been very important to scrutinize the flavor content of PDFs and the advent of new generations of neutrino experiments will open new possibilities. We want here to stress that they can help to access the elusive chiral-odd generalized parton distributions.
Some effort has already been dedicated to this question within the domain of virtual photon mediated processes \cite{transGPDacc}, but the main result is that the studies are likely to be out of the abilities of present accelerators; future electron-ion colliders \cite{EIC} may help. The only exception is the proposal \cite{pion} that pion electroproduction data are sensitive to transversity  GPDs thanks to their interplay with the chiral-odd twist 3 Distribution Amplitude (DA) of the pseudo scalar mesons.

In this paper we consider the exclusive reactions
\begin{eqnarray}
\nu_l (k)N(p_1) &\to& l^- (k')D^+ (p_D)N'(p_2) \,,\\
 \bar\nu_l (k) N(p_1) &\to& l^+ (k') D^-(p_D) N' (p_2)\,,\nonumber
\end{eqnarray}
in the kinematical domain where collinear factorization  leads to a description of the scattering amplitude 
in terms of nucleon GPDs and the $D-$meson distribution amplitude, with the hard subprocess  ($q=k'-k; Q^2 = -q^2$):
\begin{equation}
W^+(q) d \to D^+ d' ~~~~~~~~ W^-(q) u \to  D^- u'\,,
\end{equation}
described by the  handbag Feynman diagrams of Fig. 1.

\begin{figure}
\includegraphics[width=0.45\textwidth]{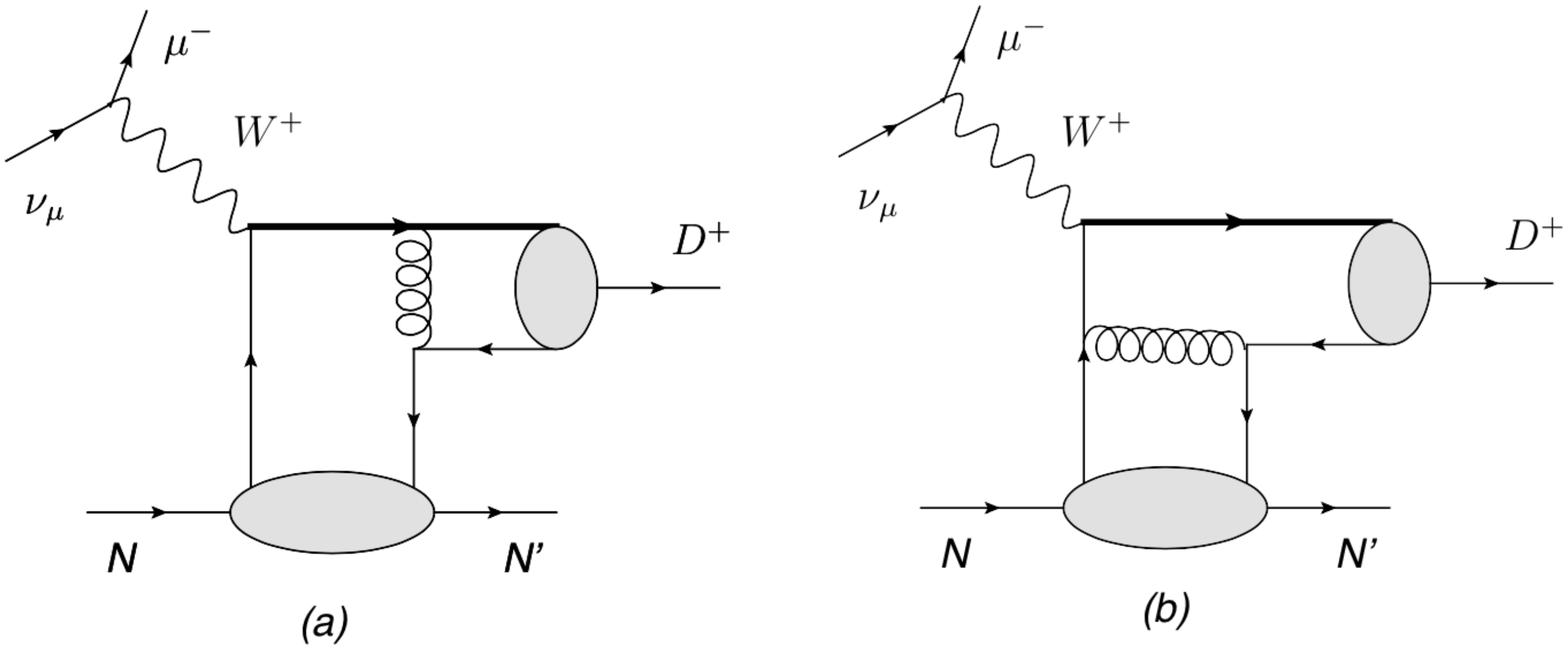}
\vspace{-1cm}
\caption{Feynman diagrams for the factorized  amplitude for the $ \nu_\mu N \to \mu^-  D^+ N'$ process; the thick line represents the heavy quark. In the Feynman gauge, diagram (a) involves convolution with both the transversity GPDs and the chiral even ones, whereas diagram (b) involves only chiral even GPDs.}
   \label{Fig1}
\end{figure}

We will demonstrate that the transverse amplitude $W_T q \to D q'$   gets its leading term in the collinear QCD framework as a convolution of chiral odd leading twist GPDs with a  coefficient function of order $\frac{m_c}{Q^2}$ (to be compared to the $O(\frac{1}{Q})$ longitudinal amplitude) and that it should be measurable in near future experiments at neutrino factories.
\paragraph*{The azimuthal dependence of  neutrinoproduction.}
The dependence of a leptoproduction  cross section on azimuthal angles is a well documented \cite{Arens} and widely used way to analyze the scattering mechanism. This procedure is helpful as soon as one can define an angle $\varphi$ between a leptonic and a hadronic plane, as for deeply virtual Compton scattering \cite{DGPR} and related processes. In the neutrino case, it reads :
\begin{eqnarray}
\label{cs}
&&\frac{d^4\sigma(\nu N\to l^- N'D)}{dx_B\, dQ^2\, dt\,  d\varphi}=\tilde\Gamma
\Bigl\{ \frac{1+ \sqrt{1-\varepsilon^2}}{2} \sigma_{- -}+\varepsilon\sigma_{00}
\\
&& +  \sqrt{\varepsilon}(\sqrt{1+\varepsilon}+\sqrt{1-\varepsilon} )(\cos\varphi\
{\rm Re}\sigma_{- 0} + \sin\varphi\
 {\rm Im}\sigma_{- 0} )\ \Bigr\},\nonumber
\end{eqnarray}
with 
\begin{equation}
\tilde \Gamma = \frac{G_F^2}{(2 \pi)^4} \frac{1}{16x_B} \frac{1}{\sqrt{ 1+4x_B^2m_N^2/Q^2}}\frac{1}{(s-m_N^2)^2} \frac{Q^2}{1-\epsilon}\,, \nonumber
\end{equation}
and the ``cross-sections'' $\sigma_{lm}=\epsilon^{* \mu}_l W_{\mu \nu} \epsilon^\nu_m$ are product of  amplitudes for the process $ W(\epsilon_l) N\to D N' $, averaged  (summed) over the initial (final) hadron polarizations.
In the anti-neutrino case,  one gets
a similar expression with $\sigma_{--} \to \sigma_{++}$ , $\sigma_{-0}\to \sigma_{+0}$, $1+ \sqrt{1-\varepsilon^2} \to 1- \sqrt{1-\varepsilon^2}$ and $\sqrt{1+\varepsilon}+\sqrt{1-\varepsilon} \to \sqrt{1+\varepsilon}-\sqrt{1-\varepsilon}$.
We use the standard notations of deep exclusive leptoproduction, namely $P=(p_1+p_2)/2$, $\Delta = p_2-p_1$, $t=\Delta^2$,  $x_B=Q^2/2p_1.q$, $y=p_1.q/p_1.k$ and $\epsilon \simeq 2(1-y)/[1+(1-y)^2]$. $p$ and $n$ are light-cone vectors ($v.n=v^+, v.p=p^-$ for any vector $v$) and $\xi=-\Delta.n/2P.n$ is the skewness variable. The azimuthal angle $\varphi$ is defined \cite{Arens}  in the  initial nucleon  rest frame as: 
\begin{equation}
sin ~\varphi = \frac {\vec q \cdot[(\vec q \times \vec p_D) \times (\vec q \times \vec k)]}{|\vec q||\vec q\times \vec p_D||
\vec q\times \vec k|}\,,
\end{equation}
while the final nucleon momentum lies in the $xz$ plane ($\Delta^y = 0$) and $\epsilon^{0123}=-1$.

We now focus on the evaluation of the longitudinal and transverse amplitudes which will (in the neutrino case)  contribute respectively to $\sigma_{00}$ and $\sigma_{--} $, while their interference will construct  $\sigma_{-0}$. Two ingredients need first to be defined, namely the  $D-$meson distribution amplitude and the transversity GPDs.

\paragraph*{$D-$meson distribution amplitude.} In the colinear factorization framework, the hadronization of the quark-antiquark pair is described by a distribution amplitude(DA) which obeys a twist expansion and evolution equations. Much work has been devoted to this subject \cite{heavyDA}.
Here, we shall restrict ourselves to a leading twist description of the $D-$meson DA, defined as (we omit the path-ordered gauge link):
\begin{eqnarray}
\langle 0 | \bar d(y) \gamma^\mu \gamma^5c(-y) |D(p_D) \rangle  =
   i f_D P^\mu \int^1_0 e^{i(2z-1) p_D.y}   \phi_D(z)\,, \nonumber
      \end{eqnarray}
 where $\int_0^1 dz ~ \phi_D(z) = 1$ and  $f_D= 0.223$ GeV.

   
   \paragraph*{Transversity GPDs.}

The twist 2 transversity GPDs have been defined \cite{transGPDdef} and their experimental access much discussed  \cite{transGPDacc}. They correspond to the tensorial Dirac structure $\bar \psi^q\, \sigma^{\mu\nu} \,\psi^q  $. In the nucleon case, there are four twist 2 transversity GPD defined as:
\begin{eqnarray}
  \label{tGPD}
\lefteqn{ \frac{1}{2} \int \frac{d z^-}{2\pi}\, e^{ix P^+ z^-}
  \langle p_2,\lambda'|\, 
     \bar{\psi}(-{\textstyle\frac{1}{2}}z)\, i \sigma^{+i}\, 
     \psi({\textstyle\frac{1}{2}}z)\, 
  \,|p_1,\lambda \rangle \Big|_{z^+= \mathbf{z}_T=0} } 
\nonumber \\
&=& \frac{1}{2P^+} \bar{u}(p_2,\lambda') \left[
 H_T^q\, i \sigma^{+i} +
 \tilde H_T^q\, \frac{P^+ \Delta^i - \Delta^+ P^i}{m_N^2} \right.
\nonumber \\
&& \left. +
  E_T^q\, \frac{\gamma^+ \Delta^i - \Delta^+ \gamma^i}{2m_N} +
 \tilde E_T^q\, \frac{\gamma^+ P^i - P^+ \gamma^i}{m_N}
  \right] u(p_1,\lambda).
\hspace{2em}
\end{eqnarray}
The leading GPD $H_T(x,\xi,t)$ is equal to the transversity PDF in the $\xi=t=0$ limit, which has recently been argued \cite{BC} to be sizable  for the $d-$quark, which is contributing to the process under study here. 

\paragraph*{The longitudinal amplitude.}
The longitudinal leading twist leading order amplitude  has been computed previously \cite{Kopeliovich:2012dr} for a pseudo scalar light meson and the calculation for the $D-$meson case is but a slight modification of this result. It is a convolution of chiral-even GPDs $H^d(x,\xi,t), \tilde H^d(x,\xi,t), E^d(x,\xi,t)$ and $\tilde E^d(x,\xi,t)$ and reads:
\begin{eqnarray}
T_{L}  =  \frac{-i C}{Q} &&\bar{N}\left(p_{2}\right) \left[\mathcal{H}_{D}\hat n+\frac{1}{2m_{N}}\mathcal{E}_{D}i\sigma^{n\Delta}\right.
 \nonumber \\
 && ~~~\left. -\tilde{\mathcal{H}}_{D}\hat n \gamma_{5} -\frac{\Delta.n}{2m_{N}}\tilde{\mathcal{E}}_{D}\gamma_{5} \right] N(p_1)\,,
\end{eqnarray}
with  $C = \frac{8\pi}{9}  \alpha_{s}V_{dc} $ and ($\bar z = 1-z$) :
\begin{eqnarray}
\mathcal{F}_{D}(\xi,t) =   f_{D}\int dz \frac{\phi_{D}(z)}{\bar z} \int dx \frac {F^d(x,\xi,t)}{x-\xi+i\epsilon},
\end{eqnarray}
for any chiral even $d-$quark  GPD in the nucleon $F^d(x,\xi,t)$;   $V_{dc}$ the CKM matrix element.

\paragraph*{The transverse amplitude up to $O(m_c/Q^2)$.}
It is straightforward to show that the transverse amplitude vanishes at the leading twist level in the zero quark mass limit. For chiral-even GPDs, this comes from the colinear kinematics appropriate to the calculation of the leading twist coefficient function; for chiral-odd GPDs, this comes from the odd number of $\gamma$ matrices in the Dirac trace. 

To estimate the transverse amplitude, one thus needs to evaluate quark mass effects. 
Indeed, it has been demonstrated \cite{Collins:1998rz} that hard-scattering factorization of meson leptoproduction \cite{fact3} is valid at leading twist with the inclusion of heavy quark masses in the hard amplitude. This proof is applicable independently of the relative sizes of the heavy quark masses and $Q$, and the size of the errors is a power of $\Lambda/ Q$ independently of the mass scale. In our case, this means including the part $\frac{m_c}{k_c^2-m_c^2}$ in the off-shell heavy quark propagator (see the Feynman graph  on Fig. 1a) present in the leading twist coefficient function. 
Adding this part of the heavy quark propagator  has no effect on the calculation of the longitudinal amplitude (because of the odd number of $\gamma$ matrices in the Dirac trace) but leads straightforwardly to a non-zero transverse amplitude when a chiral-odd transversity GPD is involved.

In the Feynman gauge, the non-vanishing $m_c-$dependent part of the Dirac trace in the hard scattering part depicted in Fig. 1a reads:
\begin{eqnarray}
Tr&[&\sigma^{pi}\gamma^\nu \hat p_D \gamma^5 \gamma_\nu \frac{m_c}{k_c^2-m_c^2+ i \epsilon}(1- \gamma^5)\hat \epsilon \frac{1}{k_g^2+ i \epsilon} ] \\
&& = \frac{2 Q^2}{\xi} \epsilon_\mu[i \epsilon^{\mu p i n} - g^{\mu i}_\perp]  \frac{m_c}{k_c^2-m_c^2+ i \epsilon} \frac{1}{k_g^2 + i \epsilon} \, , \nonumber
 \end{eqnarray} 
 where $k_c$ ($k_g$) is the heavy quark (gluon) momentum and $\epsilon$ the polarization vector of the $W-$boson (we denote $\hat p = p_\mu \gamma^\mu$ for any vector $p$).
The fermionic trace vanishes for the diagram shown on Fig. 1b thanks to the identity $\gamma^\rho \sigma^{\alpha \beta}\gamma_\rho = 0$. The denominators of the propagators 
read:
\begin{eqnarray}
&&k_c^2 - m_c^2 = \frac{Q^2 + m_c^2}{2 \xi} (x-\xi) \,,\\
 && k_g^2 = \bar z [\bar z m_c^2 + \frac{Q^2 + m_c^2}{2 \xi} (x-\xi)]\, . \nonumber
 \end{eqnarray}  
The transverse amplitude is then written  as ($\tau = 1-i2$):
\begin{eqnarray}
T_{T} & = &\frac{iC \xi m_c}{\sqrt 2 Q^2}  \bar{N}(p_{2}) \left[  {\mathcal{H}}_{T}^\phi i\sigma^{n\tau} +\tilde {\mathcal{H}}_{T}^\phi \frac{\Delta^{\tau}}{m_N^2} \right. \nonumber \\
&&  + {\mathcal E}_{T}^\phi \frac{\hat n \Delta ^{\tau}+2\xi  \gamma ^{\tau}}{2m_N} - \tilde {\mathcal E}_{T}^\phi \frac{\gamma ^{\tau}}{m_N}] N(p_{1}), 
\end{eqnarray}
in terms of  transverse form factors that we define as  :
\begin{eqnarray}
{\cal F }_T^\phi=f_{D}\int \frac{\phi(z)dz}{\bar z}\hspace{-.1cm}\int \frac{F^d_T(x,\xi,t) dx }{(x-\xi+i\epsilon) (x-\xi +\alpha \bar z+i\epsilon)},
\label{TFF}
 \end{eqnarray} 
where $F^d_T$ is any d-quark transversity GPD, $\alpha = \frac {2 \xi m_c^2}{Q^2+m_c^2}$ and we shall denote ${\bar {\mathcal{E}}_T^\phi}=\xi{\mathcal{E}}_T^\phi-{\tilde {\mathcal{E}}_T^\phi}$ .
 
 \paragraph*{Observables.}
 We now calculate from $T_{L} $ and $T_{T} $ the  quantities $\sigma_{00}$, $\sigma_{--}$ and  $\sigma_{-0}$ which enter into the observables defined by Eq.3. 
 The longitudinal cross section $\sigma_{00}$ is straightforwardly obtained by squaring the amplitude $T_{L} $; at zeroth order in $\Delta_T$, it reads  :
\begin{eqnarray}
\sigma_{00} =    \frac{C^2} {2 Q^2}\biggl\{8 (|{\mathcal{H}}^2_{D}| + |\tilde{\mathcal{H}}^2_{D}|)(1-\xi^2) + |\tilde{\mathcal{E}}^2_{D}|\frac{1+\xi^2}{1-\xi^2}\biggr\}.\,
\end{eqnarray}
At zeroth order in $\Delta_T$, $\sigma_{--}$ reads:
\begin{eqnarray}
\sigma_{--} =   \frac{4\xi^2 C^2 m_c^2}{Q^4}&\biggl\{&(1-\xi^2)|{\mathcal{H}_T^\phi}|^2  + \frac{\xi^2}{1-\xi^2} | {\bar {\mathcal{E}}_T^\phi}|^2\nonumber\\
&&-2\xi \mathcal{R}e [ \mathcal{H}_T^\phi {\bar {\mathcal{E}}_T^{\phi *}}]\biggr\} . 
\end{eqnarray}
The  interference cross section   $\sigma_{-0}$ vanishes at zeroth order in $\Delta_T$. Thus the first non-vanishing contribution being linear  in $\Delta_T/m_N$ reads (with $\lambda = \tau^* = 1+i2$):
\begin{eqnarray}
&&\sigma_{-0} = \frac{- \xi \sqrt 2 C^2}{m_N}  \frac{ m_c}{Q^3} \,
\biggl\{-i \mathcal{H}_T^{*\phi} \mathcal {\tilde E}_D \xi(1+\xi)\epsilon^{pn\Delta  \lambda } \nonumber \\
&& +   \mathcal{ H}_T^{*\phi} \Delta^\lambda [ -(1+\xi) \mathcal {E}_D]  
 +   \mathcal{\tilde H}_T^{*\phi} \Delta^\lambda [2\mathcal {H}_D -\frac{2\xi^2}{1-\xi^2} \mathcal {E}_D]  \nonumber \\
&&+  \mathcal{E}_T ^{*\phi} \Delta^\lambda [(1-\xi^2) \mathcal {H}_D  - \xi^2\mathcal {E}_D] \\
&& +  {\bar {\mathcal{E}}_T^{\phi *}}[ \Delta^\lambda [(1+\xi)\mathcal {H}_D  +\xi \mathcal {E}_D]+i (1+\xi)\epsilon^{p n \Delta \lambda } \mathcal {\tilde H}_D  ]\ \biggr\}.\nonumber 
\end{eqnarray}
 Let us finally estimate the magnitudes of the quantities  ${\cal R}e (\sigma_{- 0}) $ and ${\cal I}m (\sigma_{- 0})$ which are directly related to the observables $<cos \varphi>$ and $<sin \varphi>$ through
     \begin{eqnarray}
  <cos ~\varphi>&=&\frac{\int cos ~\varphi ~d\varphi ~d^4\sigma}{\int d\varphi ~d^4\sigma}= K_\epsilon\, \frac{{\cal R}e \sigma_{- 0}}{\sigma_{0 0}}  \,, \nonumber \\
   <sin~ \varphi>&=&K_\epsilon \frac{{\cal I}m \sigma_{- 0}}{\sigma_{0 0}}  \,,
   \end{eqnarray} 
   with $K_\epsilon =\frac{\sqrt{1+\varepsilon}+\sqrt{1-\varepsilon} }{2 \sqrt{\epsilon} }$ and where we consistently neglected the $O(\frac{m_c^2}{Q^2})$ contribution of $\sigma_{--}$ in the denominator. 
It should be noted that since Eq. 12 has a legitimate limit for small $\alpha$, the dependence on the heavy meson DA effectively factorizes in the transverse form factors ${\mathcal{H}}^\phi_T$, ${\mathcal{E}}^\phi_T$, ${ \mathcal{\tilde H}}^\phi_T$, ${ \mathcal{\tilde E}}^\phi_T$ as it does in $\mathcal{F}_D$  (Eq. 7), and thus disappears in the  ratios of Eq. 15. The complete formula obtained from Eq. 12 and Eq. 14 is quite lengthy; if we make the plausible assumption  that  ${\mathcal {\tilde H}}(\xi,t) <<{\mathcal H}(\xi,t)$ because of the known smallness of the ratio of the helicity dependent to the helicity independent $d-$quark distribution function, and moreover consider $\xi$ to be sufficiently small (say $\xi <0.3$), we get quite simple approximate results:
  \begin{eqnarray}
&& \hspace{-1cm}<cos\varphi> \approx \frac{K{\cal R}e[{\mathcal{H}}_{D} (2{\mathcal{\tilde H}}_{T}^\phi + {\mathcal{E}}_{T}^\phi + {\bar {\mathcal{E}}_T^\phi})^*- \mathcal {E}_D {\mathcal{H}}_{T}^{\phi *}]} {8|{\mathcal{H}}^2_{D}| + |\tilde{\mathcal{E}}^2_{D}|} \,, \nonumber \\
 && \hspace{-1cm}<sin\varphi> \approx  \frac{K{\cal I}m[{\mathcal{H}}_{D} (2 {\mathcal{\tilde H}}_{T}^\phi + {\mathcal{E}}_{T} ^\phi+ {\bar {\mathcal{E}}_T^\phi})^*- \mathcal {E}_D {\mathcal{H}}_{T}^{\phi *} ]}{8 |{\mathcal{H}}^2_{D}| + |\tilde{\mathcal{E}}^2_{D}|} \, ,\nonumber \\
K&=& -\frac{\sqrt{1+\varepsilon}+\sqrt{1-\varepsilon} }{2 \sqrt{\epsilon} } ~\frac{ 2\sqrt 2 \xi  m_c}{Q } \, \frac{\Delta_T}{m_N} \,.
    \end{eqnarray} 
In our kinematics, $\Delta^1=\Delta^x=\Delta_T$, $\Delta^y=0$, $\epsilon^{pn\Delta\lambda}=-i\Delta_T$.
\paragraph*{Conclusion.}
We thus have defined a new way to get access to the transversity chiral-odd generalized parton distributions, the knowledge of which would shed a new light on the quark structure of the nucleons. Our main results are
\begin{itemize}
\item Collinear QCD factorization allows to calculate neutrino production of $D-$mesons in terms of GPDs.
\item Chiral-odd and chiral-even GPDs contribute to the amplitude for different polarization states of the W (Eq.6 and Eq.10).
\item The azimuthal dependence of the cross section allows to get access to chiral-odd GPDs (Eq.3).
\item There is no small factor preventing the measurement from  being feasible, provided $\xi, \frac{m_c}{Q}, \frac{\Delta_T}{m_N}$ are not too small (Eq.14 and Eq.16).
\end{itemize}
 Planned high energy neutrino facilities \cite{NOVA} which have their scientific program oriented toward the understanding of neutrino oscillations  or elusive inert neutrinos may thus allow - without much additional equipment - some important progress in the realm of hadronic physics. We do not claim that the experimental measurement of the observables proposed in Eq.15  and Eq.16 will be an easy task. But the observables that we propose are certainly not a one per cent effect. One can check that the factor $K$ in Eq. 16 is not small   ($K = 0(\frac{.3}{\sqrt{\epsilon}})$) if we focus on $Q$ in the range of $2-3$ GeV and $\Delta_T/m_N \approx 0.5$ (this conclusion is unchanged if we include terms of order $(\Delta_T/m_N)^2$ in Eq. 12 and 14).  However, one may anticipate that the measurement of $\varphi$ will be difficult since the reconstruction of the $D-$meson will not be complete, and that the exclusivity of the reaction will not be easy to prove since a neutrino beam has a wide energy spread and the target nucleon is inside a nucleus. A dedicated feasibility study is thus obviously needed to decide whether the observables defined here can be experimentally measured in a definite experimental set-up, but this is not within the goal of the present paper. 


\paragraph*{Acknowledgements.}
\noindent
We thank O.V. Teryaev for   useful discussions.
 This work is partly supported by the Polish Grant NCN No DEC- 2011/01/B/ST2/03915 and by the French grant ANR PARTONS (Grant No. ANR-12-MONU-0008-01).

\end{document}